\documentclass[aps,12pt,floatfix,tightenlines,showkeys,showpacs]{revtex4}
\usepackage{epsfig}
\usepackage{graphicx}
\usepackage{amssymb}
\newcommand{\nn}{\nonumber\\&&}

\newcommand{\kslash}{\not\hspace{-0.7mm}k}\newcommand{\lslash}{\not\hspace{-0.7mm}l}
\newcommand{\pslash}{\not\hspace{-0.7mm}p}

\newcommand{\qslash}{\not\hspace{-0.7mm}q}

\newcommand{\pslashoff}{\not\hspace{-0.7mm}p^{\rm off}}
\newcommand{\munu}{{\mu\nu}} \newcommand{\mn}{{\mu\nu}}
\newcommand{\ben}{\begin{displaymath}}
\newcommand{\een}{\end{displaymath}}
\newcommand{\be}{\begin{equation}}
\newcommand{\ee}{\end{equation}}
\newcommand{\bea}{\begin{eqnarray}}
\newcommand{\eea}{\end{eqnarray}}

\newcommand{\eq}[1]{Eq.~(\ref{#1})}

\begin{document}

\title{\bf  \hskip10cm NT@UW-12-10\\
{ Nuclear Quasi-Elastic Electron Scattering  Limits Nucleon Off-Mass Shell Properties  
  	}}
                                                                           
\author{Gerald~A.~Miller,$^1$ Anthony~W.~Thomas$^2$,  and  Jonathan~D.~Carroll$^2$}
\affiliation{$^1$ Department of Physics, University of Washington,  Seattle, WA 98195-1560,\\
$^2$
CSSM, School of Physics and Chemistry, University of Adelaide, Adelaide SA 5005, Australia \\
}

\date{\today}

\begin{abstract}
The use of quasi-elastic electron nucleus scattering is shown to provide significant constraints
on models of the proton electromagnetic form factor of off-shell nucleons.  Such models can be constructed to be consistent with constraints from current conservation
and 
low-energy theorems, while also providing a contribution to the Lamb shift that might potentially resolve the proton radius puzzle in muonic hydrogen.
 However,  observations of quasi-elastic scattering limit the overall strength of the off-shell form factors to values that  correspond to small contributions to the Lamb shift.
 \end{abstract}
\maketitle     

\section{Introduction}

The structure of nucleons bound in the nucleus is different than that of free nucleons. A prominent example is the EMC effect in which the influence of the medium is  known to modify  the quark   distribution functions of nucleons, see the reviews~\cite{emc}. An even older example of medium effects involves the neutron which
lives forever in stable nuclei. This is because the effects of the $pe^-\nu$ component of the  free neutron wave function
are suppressed by the effects of binding energy  in nuclei.
The existence of medium effects on the structure of nucleons cannot be denied, but elucidating all of the 
possible effects and the relations between them is a task for ongoing research. We shall focus here on medium modifications of proton electromagnetic form factors~\cite{Paolone:2010qc}, which potentially affect quasi-elastic scattering~\cite{Meziani:1984is}, \cite{Mckeown:1986kn},
 and may contribute to solving the proton radius puzzle~\cite{Miller:2011yw}.

A prominent example of medium modifications is the work of Ref.~\cite{Paolone:2010qc} which involves measuring   the double ratio of proton-recoil polarization-transfer coefficients of the quasielastic $^4$He(e,e'p)$^3$H reaction with respect to the elastic $^1$H(e,e'p) reaction  which is sensitive to possible medium modifications of the proton form factor in $^4$He. Measurements  of this double ratio at four-momentum transfers squared   between 0.4 GeV$^2$ and 2.6 GeV$^2$ performed at both Mainz and Jefferson Lab find a reduction of about 10\% in the double ratio, which corresponds to a similar reduction in the ratio of electric to magnetic form factors $G_E/G_M$. Models which treat the  nucleon as a bound state of three quarks which move under the influence of quarks in other nucleons, consistent with the EMC effect and much nuclear phenomenology, predict such a reduction~\cite{Lu:1998tn}.  Alternative explanations 
%%%new here
involving final state interactions are possible (see the discussion in Ref.~\cite{Paolone:2010qc} and references therein), but seem to be incompatible with the totality of relevant data.

It is noteworthy that measurements of quasi-elastic scattering  are related to one of the most perplexing physics issues of recent times--the proton radius puzzle.  This puzzle originates in the 
  extremely precise extraction of the proton radius~\cite{pohl} from the measured 
energy difference between the $2P_{3/2}^{F=2}$ and  $2S_{1/2}^{F=1}$ states of muonic hydrogen (H).
This Lamb shift depends on the finite size of the proton's electromagnetic field. The extreme precision of the measurement
 leads to an extracted value of the  proton radius  smaller than 
the CODATA~\cite{codata} value (extracted mainly from electronic H) by about 4\% or 5.0 
standard deviations.  This implies~\cite{pohl} that either the Rydberg constant has to be 
shifted by 4.9 standard deviations or that 
present  QED calculations for hydrogen are insufficient. 
Since the Rydberg constant is extremely well measured, and the 
QED calculations seem to be very extensive and highly accurate, the muonic H finding has presented 
a significant puzzle to the entire physics community.

We need a brief discussion of the relevant phenomenology. Pohl {\it et al.} show, perturbatively that 
the energy difference
between the  $2P_{3/2}^{F=2}$ and  $2S_{1/2}^{F=1}$ states, $\Delta\widetilde{E}$   is given by
\bea 
\Delta\widetilde{E}=209.9779(49)-5.2262r_p^2+0.0347 r_p^3  \;{\rm meV},\label{rad}
\eea
where $r_p$ ($r_p^2$ is related to the slope of $G_E(Q^2)$ at $Q^2=0$) is given in units of fm. Using  
this equation, and the experimentally measured value, $\Delta\widetilde{E}=206.2949$ meV, one can  see that the difference between the Pohl and CODATA values of the proton radius  
would be   removed by an increase of the first term on the rhs of Eq.~(1) 
%%%first change
by just 0.31 meV=$3.1\times 10^{-10}$  MeV, but an effect of even half that much would
be large enough to dissipate the puzzle.  

This proton radius puzzle has been attacked from many different directions~\cite{Jaeckel:2010xx}--\cite{Carlson:2011af},\cite{Miller:2011yw}
The present communication is intended to investigate  the  hypothesis~\cite{Miller:2011yw} that  the off-mass-shell dependence of the proton electromagnetic form factor 
  that occurs in  the lepton-proton two-photon exchange interaction can account for the 0.31 meV.
 This idea is attractive because the computed effect is proportional to the lepton mass to the fourth power, and 
so is capable of being relevant for muonic atoms, but irrelevant for electronic atoms.  To make  a calculation one needs to postulate a specific dependence  of the electromagnetic form factor as a function of the  proton's virtuality (difference between 
the square of the proton's four-momentum vector and the square of the proton mass). 
Many different functional forms are possible and Ref.~\cite{Miller:2011yw}  chose one that accounted for the difference between the muonic and electronic hydrogen measurements. 

Ref.~\cite{Carlson:2011dz} uses  a dispersion analysis   of  the two-photon exchange term \cite{Pachucki:1999zza,Carlson:2011zd}
   to provide limits on the size of the allowed off-shell effect of the specific chosen form of Ref.~\cite{Miller:2011yw}.   
   %%%new here
 This is done by expressing the virtual Compton scattering amplitude implied in    Ref.~\cite{Miller:2011yw} in terms of
 the invariant $T_{1,2}$ and relating those amplitudes to electric and magnetic polarizabilities. The ones used in Ref.~\cite{Miller:2011yw}  are shown in Ref.~\cite{Carlson:2011dz} to be far larger than the experimentally measured ones. 
The accuracy  of such dispersion relation  approaches may be less than previously thought~\cite{WalkerLoud:2012bg}.  
Nevertheless, we construct a new model of the off-shell form factor that  is consistent with all of the conditions mentioned in Ref.~\cite{Carlson:2011dz}. 
These conditions are derived using the constraints of  second order in chiral perturbation theory.  Very recently Birse \& McGovern~\cite{Birse:2012eb} evaluated the constraints to  fourth-order in chiral perturbation theory.  We show how to develop off-shell models that are consistent with any  order of chiral perturbation theory.
Moreover, these kinds of models are testable in a variety of arenas, and in particular
it is of interest  to examine the consequences of the proposed off-shell model of~\cite{Miller:2011yw} for electron-nucleus scattering.

The idea we consider is that a bound nucleon can be taken evanescently off its mass-shell by virtue of its interactions with other nucleons,
and that the consequences of using the model~\cite{Miller:2011yw}  can therefore be tested. We make an explicit calculation of how the ratio of proton electromagnetic form factors $G_E/G_M$ is changed in the nuclear medium according to the model of 
Ref.~\cite{Miller:2011yw}, and confront the ensuing predictions with the data of  Ref.~\cite{Paolone:2010qc}, Sect.~II.  The model of~\cite{Miller:2011yw}   is shown
to yield medium modifications of the ratio of  $G_E/G_M$  far in excess of the observed effects. A new model is constructed in Sect.~III that leaves the ratio $G_E/G_M$ unmodified in the medium. This  model also is constructed to be consistent with the restrictions of any finite order in chiral perturbation theory. 
The off-shell modification  depends on an overall strength parameter denoted as $\lambda$, which is limited by quasi-elastic scattering. 
The corresponding change in the Lamb shift is computed in Sect.~IV. We find that the use of the  largest  values of $\lambda $ allowed by quasi-elastic scattering
lead to changes in the Lamb shift that are far too small to account for the proton radius puzzle.

\section{Off-shell proton form factor in quasi-elastic electron scattering}

The version 
of the Dirac form of the electromagnetic vertex operator for an interaction between one on-mass-shell and  one off mass-shell nucleon used in Ref.~\cite{Miller:2011yw} can be expressed as 
 \bea&&
 \Gamma_{\rm med}^\mu(p',p)=\gamma^\mu F_1(q^2)+{(p+p')^\mu\over 2M} {\pslashoff-M\over M}{  {-\lambda q^2\over b^2}\over 1-q^2/\Lambda^2}F_1(q^2),\label{gamma}\\&&=\gamma^\mu F_1(q^2)+\delta \Gamma^\mu\\&&
 \delta \Gamma^\mu\equiv {(p+p')^\mu\over 2M} {\pslashoff-M\over M}F_1(q^2)F(q^2)\\&&
 F(q^2)\equiv {{-\lambda q^2\over b^2}\over 1-q^2/\Lambda^2}
\eea
for a photon interacting with an proton that is initially off its mass shell, 
where $M$ is the nucleon mass, $p'=p+q$, either $p^\mu$ or ${p'}^\mu$ are off the mass shell and $p^{\rm off}$ is the four momentum of the off shell nucleon
($(p^{\rm off})^2\ne M^2$).
Note that this was one of the three possible forms ($\cal O_a$) of operators proposed in Ref.~\cite{Miller:2011yw}. Other forms are possible.
The values 
\bea {\lambda\over b^2}={2\over (79\;{\rm MeV})^2}, \Lambda=841 \;{\rm MeV} \label{num}\eea
were used in   \cite{Miller:2011yw} to give a contribution to the Lamb shift large enough to allow the CODATA value of the proton radius to be  consistent with the Pohl experiment. We note that the use of \eq{gamma} in   \cite{Miller:2011yw}  was consistent with current conservation. Replacing $(p+p')^\mu $ by $(p+p')^\mu -(p+p')\cdot q q^\mu/q^2 $ (where $q$ is the virtual photon momentum gives no change to the computed shift in the atomic binding energy. 

The use of the vertex function of \eq{gamma} in computing virtual-photon-proton Compton scattering leads to new contributions at low values $Q^2\equiv-q^2$. It is worthwhile to compare these effects with those of standard formulations
in which two invariant amplitudes $T_{1,2}$ appear.  Given the model of \eq{gamma} there is a new contribution to $T_2$ (but not $T_1$), which has been found to be  Ref.~\cite{Carlson:2011dz} 
\bea T_2^{\rm off}\approx -{2\lambda\over \pi M b^2}Q^2,\eea 
for small values of $Q^2$.  In standard formulations the coefficient of the $Q^2$ term of $T_2$ is given in terms of the electric $\alpha_E$ and magnetic $\beta_M$ polarizability of the proton as $Q^2/e^2 (\alpha_E+\beta_M)$. Equating the coefficients of $Q^2$ gives
$\lambda/b^2=-0.018 (2)/(79\;{\rm MeV})^2$. Thus the constraints imposed by the known electromagnetic polarizabilities yield a value of $\lambda/b^2$ that is about 55  smaller than  and of the opposite sign to the value given in 
\eq{num}~\cite{Carlson:2011dz}.  The  other models mentioned in Ref.~\cite{Miller:2011yw} were not used to compute the Lamb shift, but would correspond to different values of $\lambda/b^2$ which are of the same order of magnitude as that
of \eq{num}, and those models would therefore fare equally poorly.  Thus the model of Ref.~\cite{Miller:2011yw} is not consistent with known features of the virtual-photon-proton Compton scattering amplitude. 
However,it is worthwhile to examine the consequences of such a model for other processes to illustrate the connections
between different areas of physics. Moreover, 
 the restrictions of Ref.~\cite{Carlson:2011dz} can be removed simply by postulating that the off-shell effects of \eq{gamma} be proportional to $q^4$. 

We therefore consider quasi-elastic electron-nuclear scattering.
The basic idea is that 
a proton, bound in the nucleus is  slightly off its mass shell.  The average binding energy of a nucleon is  less than 1 percent of its mass. But even this very small binding corresponds to an off-shell effect that contradicts experiment if the
model of \eq{gamma} is used. Consider a proton bound via a Dirac mean-field Hamiltonian. The bound-state wave function $|\psi\rangle$ obeys the equation
\bea
|\psi\rangle={1\over \pslash -M}V|\psi\rangle\eea
as in Fig.~1, in which the presence of the residual nucleus is represented by the interaction $V$. 
The change in the scattering amplitude $\delta {\cal M}^\mu$ caused by the off-shell term of \eq{gamma} is given by
\bea
\delta {\cal M}^\mu=\langle\psi'|\delta \Gamma^\mu|\psi\rangle=F_1F\langle\psi'|{(p+p')^\mu\over 2M}{V\over M} |\psi\rangle,\label{amp1}\eea
in which the final state single-particle wave function is represented by $|\psi'\rangle$.

We explain the relationship between the amplitude of \eq{amp1} (inherent in Fig.~1) and the analysis of Ref.~\cite{Miller:2011yw}. There are two interaction vertices in Fig. 1, one involving the photon and one involving the strong interaction field. The intermediate virtual proton propagator appears between these two vertices. This propagator is cancelled by the 
inverse propagator in the term $\delta \Gamma$,  so that 
effectively one 
sees a contact interaction between the virtual photon, the struck proton and the residual nucleus.  
The interaction  between the  photon and the virtual proton converts  the off-shell proton into its final state, $|\psi'\rangle$.
%\end{document}
Thus the very same off-shell interaction $\delta\Gamma^\mu$ of 
%\end{document}
 Ref.~\cite{Miller:2011yw}    enters here. Here one sees the combination of $\delta \Gamma^\mu$  and
V; in the two-photon exchange term we  use a combination of  $\delta\Gamma^\mu$ and $\Gamma^\nu$.
  \begin{figure}
\epsfig{file=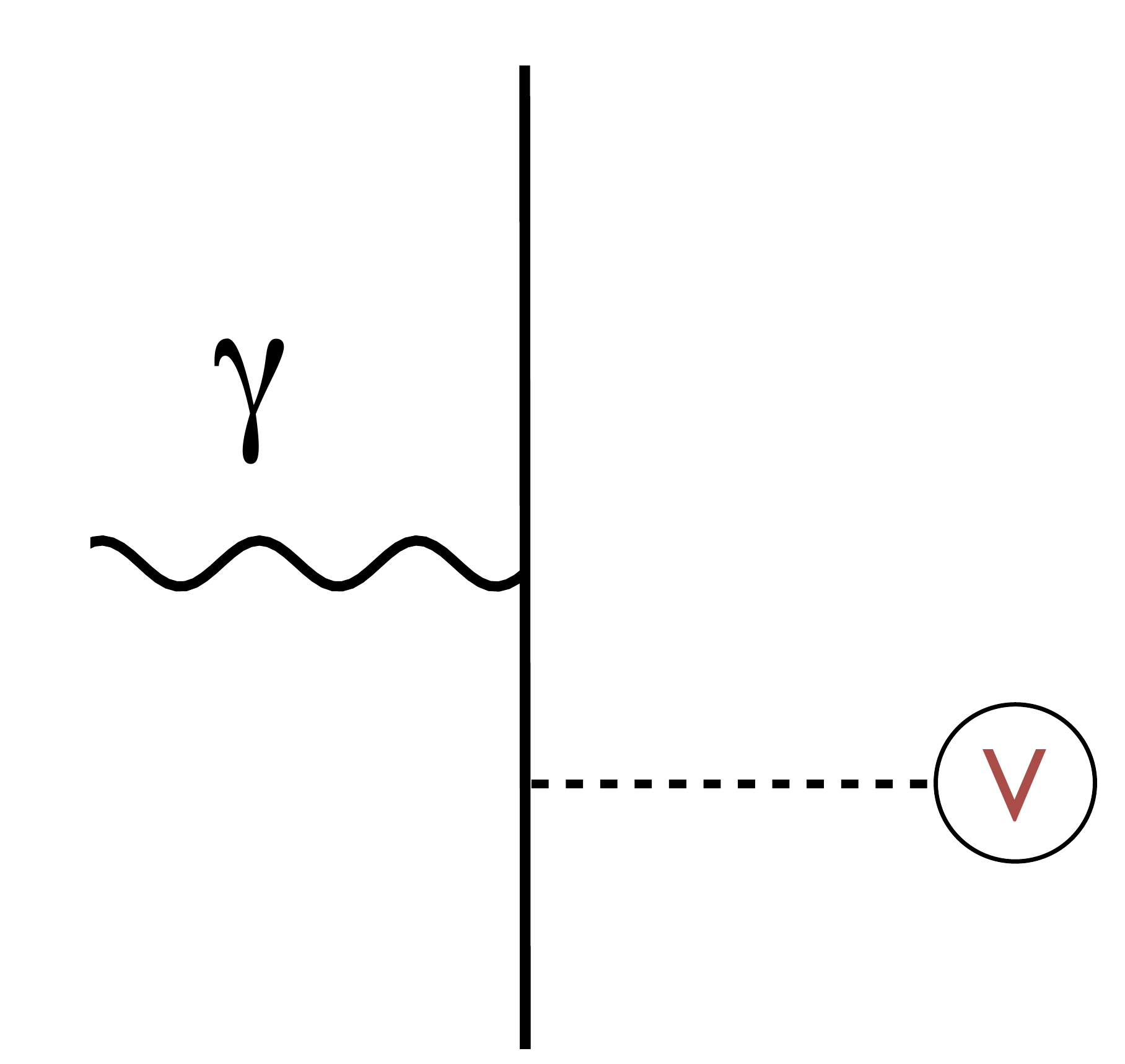, width=6.0cm}
\caption{(Color online) A photon  (wiggly line) interacts with a bound nucleon that is off its mass shell because of the 
interaction $V$.}% 
\label{fd1}\end{figure}

The evaluation of the consequences of using \eq{gamma} and \eq{amp1} proceeds by first examining the 
single particle wave functions $|\psi,\psi'\rangle$. We treat the final state wave function as a plane wave represented by a
Dirac spinor. This is reasonable for the present purpose because the effects of final state interactions are included and removed in the experimental analyses \cite{Paolone:2010qc}. We aim to  consider quasi-elastic electron scattering in the  kinematic regime in which the impulse approximation is valid, and thus use a relativistic Fermi gas model to approximate the
initial nuclear wave function. This often used approximation~\cite{Moniz:1971mt} is accurate enough for the schematic estimate that is the present aim. This is because any  correction terms are of order $V/M$ of the terms we evaluate.
Under the stated assumptions we find
\bea \delta {\cal M}^\mu
\approx F_1F\bar{u}(p'){(p+p')^\mu\over 2M}{V\over M}  u(p).%\psi_i\rangle.
 %\bar{u}(p')\Gamma_{\rm med}^\mu u(p)&=&F_1 \bar{u}(p')\left(\gamma^\mu+{(p+p')^\mu\over 2M} {\pslashoff-M\over M}{ {-%\lambda q^2\over b^2}\over1-q^2/\Lambda^2}{1\over \pslashoff-M}V \right)u(p)\nonumber\\
% &=&F_1 \bar{u}(p')\left(\gamma^\mu+{(p+p')^\mu\over 2M} {V\over M}{ {-\lambda q^2\over b^2}\over1-q^2/\Lambda^2} \right)u(p).
\label{me}
 \eea
 The largest effects of the interaction $V$ occur at the center of the  nucleus, where the density and the  mean field can be regarded as constants.  Thus we take $V$ to be number (representing an average nuclear interaction) not an operator,  define $\epsilon\equiv {V\over M} $ and use $-q^2=Q^2>0$ to obtain
 \bea
\delta {\cal M}^\mu &\approx&F_1 \bar{u}(p')\left({(p+p')^\mu\over 2M}f(Q^2) \right)u(p),\eea
 where
 \bea f(Q^2)\equiv{\epsilon {\lambda Q^2\over b^2}\over1+Q^2/\Lambda^2}.\eea
Under the stated approximations,  the present calculation is consistent with 
 current conservation.  Replacing $(p+p')^\mu $ by $(p+p')^\mu -(p+p')\cdot q q^\mu/q^2 $ gives no change to the matrix element of $\Gamma^\mu$ appearing in \eq{me}
 because the operator is evaluated between on shell spinors  so that $(p+p')\cdot q=0.$

 We gain insight by using the Gordon identity to make the replacement:
 $%\bea 
 {(p+p')^\mu\over 2M}\rightarrow \gamma^u-i{\sigma^{\mu\nu}q_\nu \over2M},\;$%\eea
so that
\bea \delta {\cal M}^\mu
&=&F_1 \bar{u}(p')\left[\gamma^\mu(1+f(Q^2))- i{\sigma^{\mu\nu}q_\nu \over2M}
f(Q^2) \right]u(p),\label{change}
\eea
which shows that 
the nuclear medium modifies  both $F_1$ and $F_2$:
\bea\delta F_1(Q^2)= F_1(Q^2) f(Q^2),\;\delta F_2(Q^2)= -F_1(Q^2)f(Q^2),\label{change1}\eea 
so that the change in $F_1$ is the negative of the  change in $F_2$.

We aim to see whether such modifications are consistent with present observations.
Strauch {\it et al.} \cite{Paolone:2010qc} measured the ratio of polarization transfer in the $^4$He nucleus to that of a nucleon  for $0.4 <Q^2<2.6$ GeV$^2$. They observed  a decrease of about  10\%.
If final state interactions are properly accounted for, this is a measurement of how the ratio  $G_E/G_M$ is changed in the medium. 
We therefore study the variation of that ratio.  Recall the definitions 
\bea
G_E=F_1-{Q^2\over 4M^2}F_2;\;G_M=F_1+F_2.\label{fdefs}\eea
The medium modified form factors ${\tilde G}_{E,M}$ are given by adding the changes in $F_{1,2}$ indicated  by \eq{change} and \eq{change1}. Note that 
${\tilde G}_M=G_M$.

The medium modified ratio is given by
\bea {{\tilde G}_E\over {\tilde G}_M}= {G_E+F_1f (1+{Q^2\over 4M^2})\over G_M}={G_E\over G_M}
\left[1+{F_1\over G_E}f (1+{Q^2\over 4M^2})\right].
\eea
We now evaluate the function $f$. 
Our aim is to see if the smallest possible values of $f$ are consistent with observations. Therefore
we take $\epsilon$ to be the ratio of the average nuclear binding divided by the nucleon mass (7 MeV for $^4$He), so $\epsilon\approx -0.007.$
Using \eq{num} we find
\bea f(Q^2)\approx  -1.8{{Q^2\over \Lambda^2}\over1+Q^2/\Lambda^2},
\eea
which ranges between -0.6 and -1.3 as $Q^2$ varies between 0.4 and 2.6 GeV$^2$.
This is between 6 and 25 times the effect  observed by  \cite{Paolone:2010qc},  if one asserts that the entire 10\% reduction of the double ratio of polarization observables  is a true 
 medium modification.  Otherwise, the discrepancy would be larger.

 One could argue that the model used to evaluate the nuclear effect, taking $V/M$ to be a  constant, is  too simple to be used. 
 %%%new here
 The interaction $V$ represents the nuclear mean field and has a spatial extent corresponding to the size of the entire nucleus. Treating this as a constant  means that we are computing form factor modifications near the center of the nucleus. This is appropriate because the experimental analyses of   \cite{Paolone:2010qc}  to which we compare includes corrections so as to approximate the situation near the center of the nucleus.
   The most evident improvement would allow $V$ to have an attractive scalar term and a repulsive vector term. Using this would  lead to a larger computed effect because the cancellation between these terms that lead to the small average binding energy of 7 MeV per nucleon would be somewhat disrupted.
Using $V/M=-0.007$ minimizes the size of the effect of using \eq{gamma} in the nuclear medium. Even with this minimization, the predicted modification of the ratio of electric to magnetic form factors is  too large.
 
 Note that the modified ratios that we compute do  not show up in full strength in the (e,e'p) experiment. This is because
 the reaction may occur at the edge of the nucleus. But such effects are far too small to account for the order of magnitude problems we encounter.
 
 The model embodied in \eq{gamma} can be regarded as ruled out by the data of  Ref.~\cite{Paolone:2010qc}.  The next section is concerned with deriving a new model.
 \section{ New Models that do not modify ratios of form factors}
 An alternate approach is to consider the Strauch data to be a constraint.  Then we have
 \bea {{\tilde G}_E\over {\tilde G}_M}\approx{G_E\over G_M},\label{c1}\eea
where the approximation means within about 10\%. We express this in terms of $F_{1,2},\delta F_{1,2},$ with $\;{\tilde F_i}=F_i+\delta F_i$ where
$\delta F_i$ being the change in $F_i$ induced by the  medium.  Using the definitions,   \eq{fdefs}, allows us to re-express the
  constraint \eq{c1} as
\bea % && {F_1+\delta F_{1}-\tau(F_2+\delta F_{2})\over F_1+\delta F_{1}+(F_2+\delta F_{2}}\approx
% {F_1 -\tau F_2\over F_1+F_2}\\&&
 %\delta F_1\;F_2=\delta F_2\;F_1\\&&
 { \delta F_1\over F_1}={\delta F_2\over F_2}.\label{c2}
 \eea
 The medium modification of the ratio $F_2/F_1$ is experimentally accessible~\cite{dd}.
 The use of \eq{c2} leads to
 \bea {F_2+\delta F_2\over F_1 +\delta F_1}={F_2\over F_1}.\label{c3}
 \eea
  
 The results \eq{c2}, \eq{c3} show why our medium modification is so large. \eq{change1} shows that $\delta F_1=-\delta F_2$. 
 
 The next step is to see if one can construct a model of off-shell form factors that satisfies the constraints of 
 \eq{c1}--\eq{c3}.
 %\section{The fix}
 This can be done if we include an effect that changes $F_2$ so that \eq{c2} is satisfied. We can do this by adding a new off-shell term of the form
 $ {\cal O}_d\equiv {\sigma^{\mu\nu}q_\nu\over 2M} (\pslash^{\rm off} -M)\cdots$. In particular, we postulate
 a new version of the off-shell vertex intended to replace the ruled-out term $\delta \Gamma^\mu$ of \eq{gamma}. Defining this operator as $ {\cal O}^\mu$,
 we try 
\bea {\cal O}^\mu=\lambda F(Q^2)[F_1(Q^2) (\gamma^\mu-{\qslash\;q^\mu\over q^2}) +i {\sigma^{\mu\nu}q_\nu\over 2M}F_2(Q^2)]\;{ (\pslash^{\rm off} -M)\over M}.\label{new}\eea
The aim is simply to find an off-shell modification that satisfies all of the constraints. 
Current conservation  is explicitly satisfied by both terms. When one calculates the diagram of Fig.~1, the
 term ${\qslash\;q^\mu/q^2}$  does not contribute because it acts between $\bar{u}(p+q)$ and $u(p)$, see \eq{me}.
With \eq{new} we have
\bea \delta F_1=\lambda F F_1,\; \delta F_2=\lambda F F_2,\eea
so that \eq{c2} is satisfied.  %It is convenient to  recall that
%rewrite  \eq{new} as
%\bea {\cal O}^\mu=\lambda F(Q^2)\Gamma^\mu { (\pslash -M)\over M},\eea
%where 
%$\Gamma^\mu$ is the conventional current operator acting between on-shell spinors.

The use of this model in the diagram of Fig.~\ref{fd1} leads to an extremely simple evaluation of the modified quasi-elastic cross section. 
The effect of the medium modification is to  simply multiply the computed quasi-elastic scattering cross section by a factor of $(1+\epsilon \lambda F(Q^2))^2$.
We thus are able to  obtain a constraint on the product $\epsilon \lambda F(Q^2)$ without specifying any of the individual factors. The form of $F(Q^2)$ is needed to compute the contribution to the Lamb shift and is discussed below.

If we assume that a 10\% change in the cross section (which is the typical uncertainty in the computation of a  cross section) can be tolerated, we find that
$  \vert\epsilon\lambda \vert F(Q^2)<0.05,$   or
\bea  \vert\lambda \vert F(Q^2)<7,\label{cons}\eea   
for $Q^2<10$ GeV$^2$. Quasi-elastic experiments have not been performed for larger values of $Q^2$. 
We take $\epsilon=-0.007$ to obtain the above constraint. 
%%% new here
We note that this number is the smallest conceivable magnitude  that one could extract from nuclear physics. A more detailed analysis would lead to a number that is larger in magnitude, and an even stronger constraint on the value of $\lambda$.
However, the limit \eq{cons} leads to a  very small contribution to the Lamb shift.

\section{ Lamb Shift Calculation}
%In our notation for $T^{\mu\nu}$ 
The invariant lepton-proton scattering amplitude arising from two photon exchange is given by
\bea {\cal M}={e^4\over (2\pi)^4}\int d^4k {L_\munu(k) T^\munu(k,p)\over (k^2+i\epsilon)^2},\label{calm}\eea
 where $p$ is the proton momentum, and is evaluated in the common rest frame.
%The factor of 1/2 arises because both lepton and hadron tensors include the crossed graph. 
The spin-averaged lepton tensor $L_\mn$ % {\bf  we need to look at the hyperfine effects of this, later}
is
given by
\bea 
&&L_\mn={1\over 4m}{\rm Tr}[(\lslash+m){\gamma_\mu (\lslash-\kslash+m)\gamma_\nu\over (k^2-2l\cdot k+i\epsilon)}],%+(\mu,\nu,l\to\nu,\mu,-k)],
\label{lmunu}\eea
where $m$ is the lepton mass. The term $T^\mn$ is the virtual photon  nucleon forward scattering amplitude. We use the definition
\bea T^{\mu\nu(k,p)}=-(g^{\mu\nu}-{k^\mu k^\nu\over k^2})T_1(k^0,k^2)+{1\over M^2}(p^\mu-{p\cdot k\over k^2}k^\mu)(p^\nu-{p\cdot k\over k^2}k^\nu)T_2(k^0,k^2),\label{tmnu}\eea
with $k^0\equiv k\cdot p /M$.
Then the use of \eq{lmunu} in \eq{calm} leads to the result:
%%here
\bea  &&{\cal M} ={2m \;e^4\over(2\pi)^4}\int {d^4k\over (k^4-(2l\cdot k)^2)(k^2+i\epsilon)^2} [- (2k_0^2+k^2) T_1
+(k^2-k_0^2)T_2]
\label{calm1}\eea

 We shall be concerned with the
change in $T^\mn$ caused by off-shell form factors, and denote
the corresponding contribution to the Lamb shift, $\Delta E_{\rm off}$.
We use the standard procedure in which the zero-energy, constant amplitude ${\cal M}$ is treated as a delta function at the origin in coordinate space as so that  %given from previous notes as 
\bea \Delta E_{\rm off}=-i{\cal M}_{\rm off}{(m_r\alpha)^3\over 8\pi},\label{shift}\eea
where the factor appearing to the right of ${\cal M}_{\rm off}$ is the square of the 2S muonic hydrogen wave function at the origin.
The   change in Compton scattering by our postulated off-shell effects is obtained by computing Compton scattering in the Born approximation. Define the conventional electromagnetic vertex operator for the absorption of a photon of momentum $k$  as $\Gamma^\mu(k)$.
Then 
\bea&& T^{\mu\nu}=T_{\rm on}^{\mu\nu}+T_{\rm off}^{\mu\nu}=\\
&&= (\Gamma^\mu(-k )+{\cal O}^\mu(-k )){1\over (\pslash+\kslash-M)} (\Gamma^\nu(k )+{\cal O}^\nu(k )) +[\mu\to\nu,\nu\to\mu,k \to-k ].\label{t1}\\&&
= T_{\rm on}^{\mu\nu} +
(\Gamma^\mu(-k ){\cal O}^\nu(k )+{\cal O}^\mu(-k )\Gamma^\nu(k )+\Gamma^\nu(k ){\cal O}^\mu(-k )+{\cal O}^\nu(k )\Gamma^\mu(-k ))\nonumber\\&&+[{\cal O}^\mu(-k )(\pslash+\kslash+M){\cal O}^\nu(k )+{\cal O}^\nu(k )(\pslash-\kslash+M){\cal O}^\mu(-k )]
\eea
%{\bf Comment --} This $T^\mn$ is $-4\pi$ times that of CVDH, but this form is correct within our formalism.
We need   the spin average, obtained  by multiplying the above by $(\pslash +M)/(4M)$ and taking the trace. In the following  text all quantities $X^\mn$  are spin averaged. 

We proceed by 
breaking  up $T^{\mu\nu}$ as a sum of  three terms:
\bea&&T^{\mu\nu}=T_{\rm on}^{\mu\nu}+\delta T^{\mu\nu}+\delta Z^{\mu\nu}
,\eea
respectively of order $\lambda^0,\;\lambda^1$ and $\lambda^2$.
Then we find
\bea&& T_{1\rm on}=-\frac{ {F_1}^2 (2p\cdot k)^2+2  {F_1} {F_2} k^4+{F_2}^2 k^4}{M \left(k^4-(2p\cdot k )^2\right)}\\
&&T_{2\rm on}%\frac{4 M Q^2 \left(4 {G_E}^2 M^2+{G_M}^2 Q^2\right)}{\left(4 M^2+Q^2\right) \left((2p\cdot q)^2-Q^4\right)}
%\\&&
=\frac{-4 F_1^2 M^2 k^2+F_2^2 k^4}{M( (2p\cdot k)^2- k^4)},\eea
which are standard results, and 
\bea
&&
\delta T^{\mu\nu}\equiv Tr\left[{(\pslash+M)\over4M}
(\Gamma^\mu(-k ){\cal O}^\nu(k )+{\cal O}^\mu(-k )\Gamma^\nu(k )+\Gamma^\nu(k ){\cal O}^\mu(-k )+{\cal O}^\nu(k )\Gamma^\mu(-k ))\right].\nn\eea
%In the above expression, the term ${\cal O}^\nu(q)$ may be replaced by ${\Gamma}^\nu(q)$. 

We find
\bea \delta T_1={-\lambda F(-k^2)\over M}\frac{F_2^2 k ^2+ 4F_1^2 M^2}{ M^2},\;\delta T_2=0,
\label{dt12}\eea
%where $Q^2\equiv-q^2$
%The relative sign of the $F_1$ and $F_2$ terms is unfortunate.
The second-order terms are obtained to be 
\bea &&\delta Z^{\mu\nu}=
Tr\left[{(\pslash+M)\over4M}\left({\cal O}^\mu(-k )(\pslash+\kslash+M){\cal O}^\nu(k )+{\cal O}^\nu(k )(\pslash-\kslash+M){\cal O}^\mu(-k )\right)\right]
\\&&\delta Z_1=\lambda^2F^2\frac{F_2 \left(F_2 \left(k_0^2-k ^2\right)-2 F_1 k^2\right)}{M^3},\;\delta Z_2=\lambda^2F^2
{(4 F_1^2 M^2- F_2^2 k^2)\over M^3}.\label{dz12}
\eea

The low-energy theorem and constraints of chiral perturbation theory constrain  $T_i(\nu,Q^2)$ for small values of $\nu$ and $Q^2$. Those constraints, as applied in Ref.~\cite{Carlson:2011dz} and earlier works,  are not modified  if we choose
$F(-k^2)\sim k^4$ for small values of $k^2$. Thus we use 
\bea F(-k^2)= {(k^2/\Lambda^2)^2\over (1+(-k^2)/\Lambda^2))^2.}\label{34}
\eea
 Birse \& McGovern~\cite{Birse:2012eb}  have provided constraints to  fourth-order in chiral perturbation theory. In general, one can satisfy the constraints to n'th order by using a more general version of $F(-k^2),\; F_n(-k^2)$:
 \bea F_n(-k^2) = {(-k^2/\Lambda^2)^n\over (1+(-k^2)/\Lambda^2))^n}.\label{35}
\eea

Now  evaluate the  integral by Wick rotation
\bea k_0\to iK_0,\;\vec{k}\to \vec{K},\;k^2\to -K_0^2-\vec{K}^2=-K^2,\;K_0=K\cos\psi,\;|\vec{K}|=K\sin\psi\eea
Integrate on $\psi$ from 0 to $\pi$,
\bea \int d^4k\cdots\to 4\pi i\int dK\; K^3\int_0^\pi d\psi \sin^2\psi\cdots\eea 
Use $e^2=4\pi\alpha$ so 
\bea 
{\cal M}=i{(4\pi\alpha)^2\over (2\pi)^4}8m\pi\int dK\; K\int_0^\pi d\psi{\sin^2\psi\over K^4+4m^2K^2\cos^2\psi}[T_1(2\cos^2\psi+1)-T_2\sin^2\psi)]\eea

Now use \eq{dt12} and \eq{dz12} in the above to get the off-shell correction.
We need
\bea &&\delta T_1+\delta Z_1={-\lambda F(K^2)\over M}\frac{- F_2^2 K^2+4 F_1^2 M^2}{ M^2}+\lambda^2F^2\frac{F_2 \left(F_2 \left(K^2\sin^2\psi\right)+2 F_1 K^2\right)}{M^3}\nn%=
%{\lambda F(Q^2)\over M^3}\left(F_1^2(-8M^2)+F_2^2(2K^2+\lambda F K^2\sin^2\psi)+2\lambda F F_1F_2K^2\right)\\&&
\delta T_2+\delta Z_2=\delta Z_2=\lambda^2F^2
{(4 F_1^2 M^2+ F_2^2 K^2)\over M^3}
\eea
\bea&&\delta {\cal M}_{\rm off}=i{8\alpha^2\over \pi}m\int dK\; K\int_0^\pi d\psi{\sin^2\psi\over K^4+4m^2K^2\cos^2\psi}[(\delta T_1+\delta Z_1)(2\cos^2\psi+1)+\delta Z_2\sin^2\psi)]\nn\eea

\begin{figure}
\epsfig{file=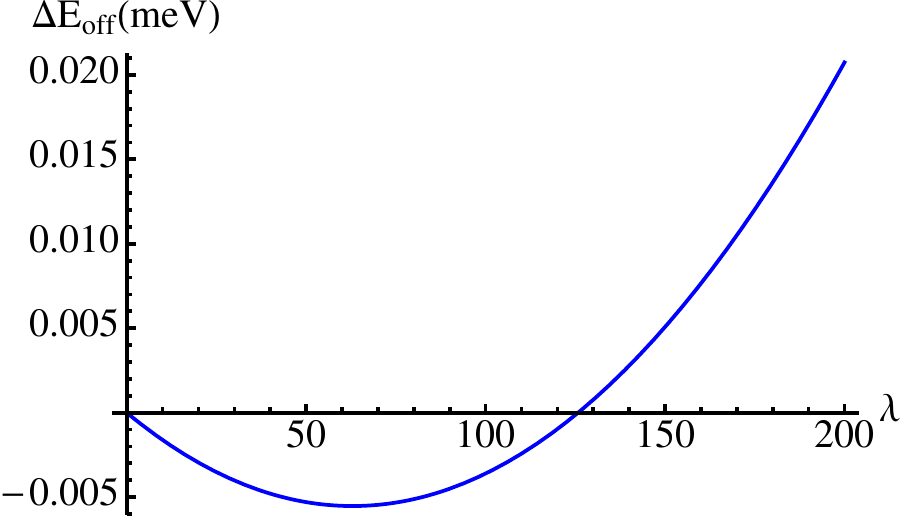, width=6.0cm}
\caption{(Color online) The energy shift $\Delta E_{\rm off}$ as a function of the parameter $\lambda$, using \eq{34}.
}% 
\label{lam}\end{figure}

%The energy shift 
The above result, along with \eq{shift}, determines the value of the computed energy shift arising from the off-shell effect.  The principal parameter is $\lambda$,   constrained to be less than about 7  (\eq{cons}) from quasi-elastic scattering data. The proton electromagnetic form factors $F_{1,2}$ are taken as dipole forms
with $\Lambda=0.841 $ GeV, and $F_2(0)=1.79$.   We start by using \eq{34} and display numerical results for values of $\lambda$ between 0 and 200 are shown in Fig.~\ref{lam}.
With $\lambda=7$, we obtain a shift of -0.001 meV, which is about 100 times too small to significantly affect the Lamb shift calculations. 
 Increasing the value of $\lambda$ provides a maximal shift of -0.005 meV, but further increases leads to a positive shift in the energy, due to the dominance of the second order terms $\delta Z_{1,2}$  for large values of $\lambda$. A positive shift in energy is of the wrong    sign to explain the proton radius puzzle.

The requirements of Birse \& McGovern~\cite{Birse:2012eb} can be satisfied by using \eq{35} with $n=3$.  The use of such a function in calculations of the Lamb shift requires even larger values of $\lambda$ to explain the proton radius puzzle. %%%new here
 The use of our limit $\lambda=7$ leads again to a very small  increase of the  Lamb shift: 0.001 meV.

\section{Discussion}
The principal result we have is that quasi-elastic electron scattering places significant limits on the off-shell dependence of the nucleon electromagnetic
vertex function. While it is possible to construct gauge-invariant  models of the off-shell behavior that are consistent with known features of the virtual photon-proton Compton scattering amplitude, these models are incapable of resolving the proton radius puzzle without causing dramatic effects in nuclear quasi-elastic scattering in   disagreement with observed data.

\section*{\it Acknowledgments:}
This research was supported by the United States Department of Energy, (GAM) grant FG02-97ER41014;   and by the Australian Research Council and the University of Adelaide (AWT, JDC) and  (JDC, in part) contract DE-AC05-06OR23177 (under which Jefferson Science Associates, LLC, operates Jefferson
Lab). GAM  gratefully acknowledges the support and gracious hospitality of the University of Adelaide while this collaboration was formed. We thank J.~Rafelski for many interesting discussions.

\end{document}